# Interstitial Solute Segregation at Triple Junctions: Implications for the Hydrogen Storage Properties of Nanomaterials

Nutth Tuchinda[1], Malik Wagih[1,2], and Christopher A. Schuh[1,3*]

[1] Department of Materials Science and Engineering, Massachusetts Institute of Technology,
77 Massachusetts Avenue, Cambridge, MA, 02139, USA

[2] Materials Science Division, Lawrence Livermore National Laboratory,
7000 East Avenue, Livermore, CA 94550, USA

[3] Department of Materials Science and Engineering, Northwestern University,
2145 Sheridan Road, Evanston, IL 60208, USA

*Correspondence to: schuh@northwestern.edu

**ABSTRACT**

At very fine grain sizes, grain boundary segregation can deviate from conventional behavior due to triple junction effects. While this issue has been addressed in prior work for substitutional alloys, here we develop a framework that accounts for interstitial sites in the grains, grain boundaries, and triple junctions of model Pd(H) polycrystals. This approach allows computation of interstitial segregation spectra separately at both defect types, which permits an understanding of segregation at all grain sizes via a size-scaling spectral isotherm. The size dependencies of dilute Pd(H) are found to be influenced not only by the triple junction content, but also by grain size-dependent lattice strains; the latter effect is evidenced by size dependencies of individual grain boundary and junction subspectra. The framework proposed here is applicable to interstitial alloys in general, and may serve as a basis for interfacial engineering in interstitial nanocrystalline alloys. As an example, we show using the dilute limit isotherm that hydrogen density can triple in nanocrystalline vis-à-vis microcrystalline Pd due to hydrogen adsorption at intergranular defect sites.

**Keywords**: Grain Boundary, Triple junction, Segregation, Nanocrystalline, Thermodynamics, Atomistic Simulation

## I. INTRODUCTION

At the finest nanocrystalline grain sizes, and as grain size decreases, the finite volume fraction of the intergranular regions can result in size-dependent thermodynamics [1–6] which can have negative (e.g., GB embrittlement [7–9]), or positive (e.g., stabilization [10–12]) consequences. Such size scaling has been explored primarily for unary materials and for substitutional alloys [13–19]. In principle, such size effects are expected for interstitial solute-defect interactions as well [20–24]. The interstitial problem is especially interesting for a number of contemporary problems: many nanocrystalline processing routes involve pickup of interstitials such as B, C, N, and O [25–29], which has implications for all of the boundary-dominated properties from stability to strength and brittleness. Conversely, intergranular regions can be potentially useful to trap, store, or transport, e.g., interstitial hydrogen for energy applications [30–33], or helium for fusion applications [34,35].

The segregation of interstitials to grain boundaries (GB) has been recently developed within a full spectral model [20,21,23]. The energy change when a solute atom occupying a bulk interstitial site moves to a site type 'i' in the intergranular regions of an FCC alloy, is shown to follow a bimodal Gaussian distribution [20] for Pd(H) with the density function ($F_i^{IG}$) of:

$$F_i^{IG} = w_{tet} \cdot \frac{1}{\sqrt{2\pi}\,\sigma_{tet}} \cdot \exp\left[-\frac{1}{2}\left(\frac{\Delta E_i^{seg} - \mu_{tet}}{\sigma_{tet}}\right)^2\right] + w_{oct} \cdot \frac{1}{\sqrt{2\pi}\,\sigma_{oct}} \cdot \exp\left[-\frac{1}{2}\left(\frac{\Delta E_i^{seg} - \mu_{oct}}{\sigma_{oct}}\right)^2\right] \quad (1)$$

where $\Delta E_i^{seg}$ is defined as the energetic change upon solute segregating from a reference bulk site to a designated site 'i'. This expression acknowledges that there are two modes of interstitial segregation in FCC metals (denoted with subscripts tet and oct), namely semi-tetrahedral and semi-octahedral sites; these two modes persist in the intergranular regions, but exhibit substantial spreading due to the atomic disorder in those regions. The spreading of the energy states is captured by the Gaussian terms of Eq. (1), where $\mu$ and $\sigma$ indicate the mean and width of each mode. The two modes contribute independently, weighted by the factors $w_{tet}$ and $w_{oct}$ reflecting the densities of those two modes.

Segregation energetics following the form Eq. (1) can lead to grain size dependence when the distribution of site types is itself a function of grain size, and thus the distribution parameters $w$, $\mu$ and $\sigma$ are too. Grain size-dependent segregation has been analyzed for substitutional alloys [13,14,16,36–38] because there are unique subdistributions of sites for defects of different order, i.e., GBs vis-à-vis triple junctions (TJs). We are not aware of any such analysis for interstitial alloys, where size dependencies should result from similar physics, but there are two unique subdistributions for the octahedral and tetrahedral sites, which complicates the way solute partitions amongst all the defect and subsite spectra. Therefore in this work we conduct a scaling study using self-similar nanocrystalline Pd grain structures, with H in solution, as a model system for the more general problem [39–41]. The solute energy distributions we quantify can be further incorporated into a spectral interstitial thermodynamic model to estimate size effects of intergranular solute chemistry.

II. SPECTRAL SEGREGATION ISOTHERM IN INTERSTITIAL NANOCRYSTALLINE ALLOYS

For the case of Pd(H), we will reference all energetics to the bulk octahedral sites, which has been demonstrated to be the preferential site type in the bulk Pd(H) system studied here [20]. In the dilute limit, the probability or concentration for interstitial solute to occupy a GB site type 'i' is then [20]:

$$X_i^{IG} = 1 + \frac{1 - X_O^C}{X_O^C} \exp\left(\frac{\Delta E_i^{seg}}{k_B T}\right) \quad (2)$$

where $X_i^{IG}$ and $X_O^C$ are the local solute concentrations at site type 'i' and bulk octahedral sites (to which we are referencing the energetics), respectively, and $k_B$ and $T$ denote the Boltzmann constant and temperature. The segregation energy is also a difference between the same two sites, i.e., $\Delta E_i^{seg} = E_i - E_O^C$.

In closed polycrystalline FCC systems, the total solute content $X^{tot}$ is conserved and can be partitioned across multiple site types via the constraint:

$$X^{tot} = (1 - f^{GB} - f^{TJ})X^C + f^{GB}\overline{X}^{GB} + f^{TJ}\overline{X}^{TJ} \tag{3}$$

with $f^{GB}$ and $f^{TJ}$ as the GB and TJ site fraction in a polycrystalline system, and $\overline{X}^{GB}$ and $\overline{X}^{TJ}$ the solute concentrations at GB and TJ. The $X^C$ here denotes the concentration of the crystalline or bulk sites in the FCC lattice, which can be further partitioned into 1/3 octahedral and 2/3 tetrahedral lattice sites:

$$X^C = \frac{1}{3}X_O^C + \frac{2}{3}X_T^C \tag{4}$$

where $X_T^C$ is the bulk tetrahedral solute concentration. The concentration of the various subpopulations can then be assessed with an isotherm, e.g., tetrahedral bulk sites can then be treated as:

$$X_T^C = 1 + \frac{1 - X_O^C}{X_O^C} \exp\left(\frac{\Delta E_{O \to T}^H}{k_B T}\right) \tag{5}$$

with $\Delta E_{O \to T}^H$ defined as the energetic penalty of a H solute atom occupying tetrahedral site instead of an octahedral one. Note that while bulk tetrahedral sites are not energetically favorable, the magnitude of $\Delta E_{O \to T}^H$ in Pd(H) is low, ~0.12 eV, and thus a non-negligible amount of H can be thermally excited into occupying this site type at finite temperatures of order hundreds of K [20] (0.12 eV/$k_B$ ≈ 1400 K). Lastly, the averaged defect concentrations $\overline{X}^{GB}$ and $\overline{X}^{TJ}$ in Eq. (3) can then be calculated by integrating Eq. (2) over the bimodal distribution of Eq. (1):

$$\overline{X}^s = \int_{-\infty}^{\infty} F_i^s \cdot \left[1 + \frac{1 - X_O^C}{X_O^C} \exp\left(\frac{\Delta E_i^{seg}}{k_B T}\right)\right]^{-1} d\Delta E_i^{seg} \tag{6}$$

where $F_i^s$ of both s = GB and s = TJ have their own characteristic distribution parameters. Combining Eqs. (3-6), this results in the total solute conservation constraint of:

$$\begin{aligned} X^{tot} = (1 - f^{IG}) \cdot &\left[\frac{1}{3}X_O^C + \frac{2}{3} \cdot \left[1 + \frac{1 - X_O^C}{X_O^C} \exp\left(\frac{\Delta E_{O \to T}^H}{k_B T}\right)\right]\right] + \\ f^{GB} \cdot &\int_{-\infty}^{\infty} F_i^{GB} \cdot \left[1 + \frac{1 - X_O^C}{X_O^C} \exp\left(\frac{\Delta E_i^{seg}}{k_B T}\right)\right]^{-1} d\Delta E_i^{seg} + \\ f^{TJ} \cdot &\int_{-\infty}^{\infty} F_i^{TJ} \cdot \left[1 + \frac{1 - X_O^C}{X_O^C} \exp\left(\frac{\Delta E_i^{seg}}{k_B T}\right)\right]^{-1} d\Delta E_i^{seg} \end{aligned} \tag{7}$$

Here the defects contribute to the total solute content via their respective defect site-type density functions ($F_i^{GB}$ and $F_i^{TJ}$) and total volume fractions ($f^{GB}$ and $f^{TJ}$), which encapsulate grain size dependencies. Thus, before proceeding to the evaluation the segregation subspectra of Pd(H) GBs and TJs at various grain sizes, we first need to quantify the geometrical scaling of interstitial site densities, which will be addressed in the next section.

## III. INTERSTITIAL SITE AND INTERGRANULAR NETWORK GEOMETRY

To establish scaling laws for interstitial solute segregation, we construct polycrystalline systems spanning a range of grain sizes, over a significant range of individual defect fractions. Self-similar polycrystals with a grain size range of d = 5.7 – 10.4 nm have been shown in prior work to effectively cover the range of defect fractions needed to convincingly separate TJ from GB phenomena [3]. Such structures are generated with the procedures described in Refs. [13,14] via Large-scale Atomic/Molecular Massively Parallel Simulator (LAMMPS) [42–46] and the software packages in Refs. [47–56] with a cutoff radius of 5.2 Å for the TJ identification algorithm of Pd sites (following Ref. [14]). Voronoi analysis [48,57,58] is conducted to identify interstitial solute sites [20,59], where H atoms within 1-Å vicinity of one another are deleted to prevent site duplication. The energetics study in Ref. [20] found that for this Pd(H) system, IG interstitial sites can be defined as the sites within a distance of 2.7 Å to the nearest intergranular solvent atom. We have adopted this definition, and apply it to the TJ interstitial site search radius: If the interstitial IG site of interest has a (solvent) Pd TJ site (i.e., a solvent atom with neighboring solvent atoms in two other grains) in the vicinity of a 2.7 Å sphere, the site is classified as a TJ site. The rest of the interstitial IG sites are then labeled as GB interstitial sites. We show the polycrystal orientation seed in Fig. 1(a) where the solvent (Pd) and interstitial (H) sites are colorized based on the defect type. The seed orientations in Fig. 1(a) are then used for the scaling of the polycrystals, resulting in the structures in Fig. 1b.

For the substitutional sites, prior work has established how a generalized polynomial form can be used to describe the defect fractions [1,13]:

$$f^s = A_3^s \left(\frac{\alpha}{d}\right)^3 + A_2^s \left(\frac{\alpha}{d}\right)^2 + A_1^s \left(\frac{\alpha}{d}\right) \tag{8}$$

$$d = 2\frac{L}{(N\frac{4}{3}\pi)^{1/3}} \tag{9}$$

where d represents the spherical equivalent grain size and α is the average characteristic grain boundary width. Each defect type (s = GB or TJ) has its own characteristic scaling parameters $A_j^s$ which depends on the grain geometry. Here we show that such polynomials are applicable for interstitial systems as well; we compute the fraction of interstitial sites at each defect type directly from the simulated structures, and present them in Fig. 2 as a function of grain size. These results are well-fitted with Eq. (8) (solid lines) with their coefficients listed in Table I. Such fitting delivers a GB characteristic length (roughly a GB thickness), α, of 1.72 nm, which is significantly larger than typical (~0.5-1 nm [3,14]) due to both the definition of interstitial IG site and the interaction of interstitials with intergranular defects which is longer-range than the substitutional one [20,23]. The nano-sized polycrystals shown here allow us to evaluate segregation energy distributions via molecular statics for ~200,000 sites at d = 5.8 nm to ~$10^6$ sites at d = 10.4 nm, which spans a reasonable coverage of TJ fraction per Fig. 2 (from $f^{GB}$ of the same scale as $f^{TJ}$, to the case where triple junctions become a small contribution). We then proceed to evaluate characteristic segregation energy for use with the density function of Eq. (1).

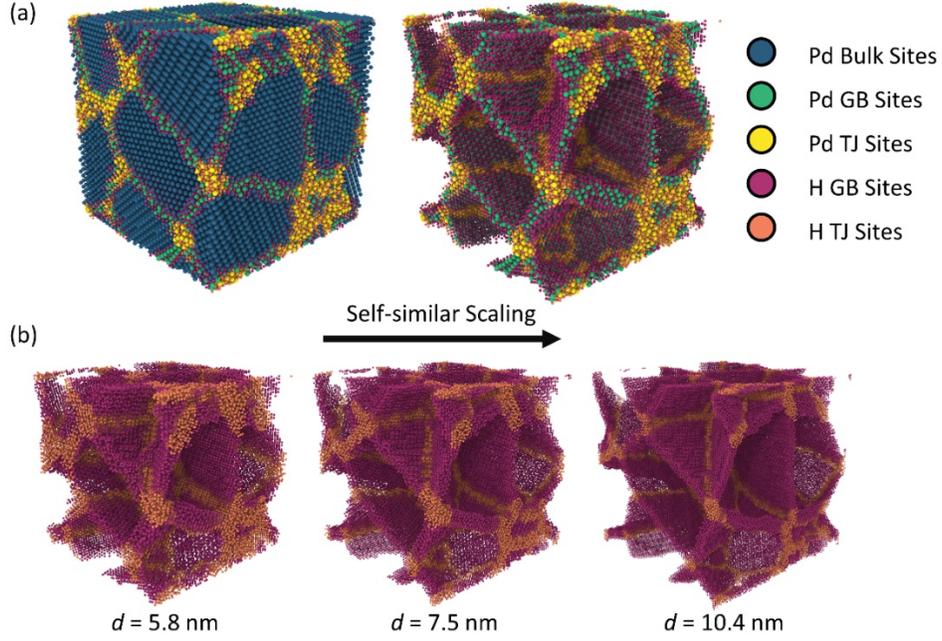

FIG. 1. A Pd polycrystal is shown in (a) for the full system with identified solvent and solute by site type. The solvent atoms are removed in (b) to visualize the triple junction network identified by our algorithm. The polycrystals are self-similarly scaled, producing crystals with varying sizes and controlled crystallography in (c). See Fig. 2 below for the coverage of defect fractions within this grain size range.

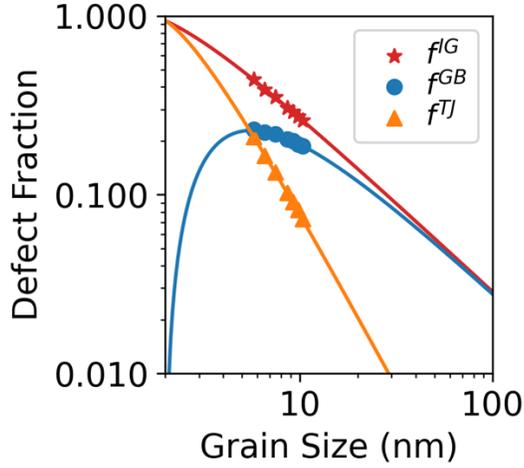

FIG. 2. interstitial defect site fraction fitted via polynomial of Eq. (8) with the coefficients listed in Table I.

TABLE I. Polynomial Coefficients for Interstitial Site Fractions of Eq. (8)

| Defect type (s) | $\alpha$ (nm) | $A_3$ | $A_2$ | $A_1$ |
| --- | --- | --- | --- | --- |
| GB | 1.72 | 1.96 | -3.64 | 1.68 |
| TJ | | -1.94 | 2.94 | 0 |

## IV. SEGREGATION ENERGY SPECTRA AT GRAIN BOUNDARIES AND TRIPLE JUNCTIONS

First, we calculate the segregation energies via fast inertial relaxation engine (FIRE) minimization [60,61], with the maximum force norm tolerance of $10^{-8}$ eV/Å. Here, the reference bulk crystalline sites are chosen from interstitial sites around 25 randomly selected Pd sites that are more than 2 nm away from GBs. The segregation energy ($\Delta E_i^{seg}$) is then calculated via:

$$\Delta E_i^{seg} = \left(E_i^{solute} - E^{poly}\right) - \left(E_{bulk}^{solute} - E^{poly}\right) \quad (10)$$

where $E_i^{solute}$ and $E_{bulk}^{solute}$ are the relaxed system energies when a H solute is occupying an IG site type 'i' and bulk octahedral site respectively, and $E^{poly}$ is the energy of a pure solvent polycrystalline system.

We first plot the total segregation energy spectra (inclusive of both GB and TJ sites) for Pd(H) at various grain sizes in Fig. 3a. The scaling shows a rather surprising behavior: these spectra do not have a discernable size dependence, despite the fact that the relative importance of GB and TJ sites is nearly inverting between the largest and smallest of the grain sizes studied (cf. Fig. 2). Over a similar range of sizes, substitutional segregation can shift rather significantly, as evidenced by prior work on several Al-based systems in Refs. [13,14]. Given the larger apparent "thickness" of the IG defects in the view of interstitials, one might expect an even larger size effect for this case.

The surprising lack of a global grain size effect therefore requires some further investigation. We plot individual subspectra in Fig. 3b where the characteristic GB and TJ subspectra are normalized by the defect volume fractions so as to focus on the relative shape changes between them. The GB and TJ spectra are indeed separable, with distinct contributions from the semi-tetrahedral and semi-octahedral sites ($w_{tet}$ and $w_{oct}$). The fitted spectral parameters for all system sizes are listed in Table II. Perhaps most importantly, the TJ sites are the most negative; all μ in Table II are shifted by ~3 kJ/mol from those of the GBs, a shift that is consistent across the subdistributions (also note that the subdistribution weighting factors w and widths σ differ for TJs). While this contrast is rather strong, there are also other observable size dependencies of the GB and TJ spectra themselves. Unlike substitutional systems such as those in Ref. [13,14], where the GB and TJ segregation spectra are found to be approximately size-independent, here we see that in interstitial systems these individual defect spectra shift (slightly, but perceptibly) with grain size. Importantly, this amounts to positive energetic shifts (i.e., in the direction of less segregation) at the finest grain sizes calculated. The degree of this shift is on the order of kJ/mol, and to first order, it appears to affect the entire distribution roughly uniformly, for instance in Table II, the GB subspectra means shift about 1 kJ/mol while maintaining roughly a size independent width.

The results in Fig. 3 are thus doubly surprising; the total interstitial segregation spectrum for intergranular segregation does not shift noticeably with grain size (whereas for some substitutional systems with a GB-TJ energetic contrast close to that in TABLE II it does, due to the increasing prominence of TJs at finer grain sizes), while the spectra for the individual defects (which normally are size-independent in substitutional segregation) are size dependent in interstitial segregation. What is more, the unexpected grain size dependence of the defect spectra in Fig. 3b provides an explanation for the unexpected grain size independence of the overall IG segregation spectra in Fig. 3a. As grain size shrinks, there is a stronger contribution from TJ sites, and those sites are indeed significantly more attractive sites for interstitial H. However, over the same grain size change, all the IG sites drift rightward in Fig 3b, opposing segregation in general. These two effects are counteracting in the present system, although we note that this may be merely fortuitous in Pd(H), and may or may not hold for other interstitial systems; this result thus calls for future work beyond our present scope to explore other such systems.

The above observation shifts the need for physical understanding to the size dependence of the individual defect segregation spectra in Fig. 3b; apparently the same set of defects, with the same crystallographies and geometrical arrangements, have different propensities for H segregation when assembled into grains of different sizes. One possibility for this unexpected result lies in the internal stress in nanocrystalline systems. Due to the interfacial tension of the GBs, finer grains are expected to adopt a state of dilation [62–64], i.e., negative pressure. Such internal stress may cause the bulk sites to expand, resulting in more energetically favorable sites vis-à-vis a large grain counterpart. This stress may also alter the site environment in GBs and TJs, making them less preferable to H solute. Before we probe into the origin of GB-TJ energetic contrast, we thus first proceed with evaluating the bulk contribution ($E_{bulk}^{solute} - E^{poly}$) in Eq. (10)) for all grain sizes calculated here.

We quantify lattice dilation by calculating interatomic distances from the polyhedral template matching algorithm [65] and plot averaged bond dilation in Fig. 4 for all grain sizes; the lattice dilation in Fig. 4a is of order 0.1% strain which is close to the order of magnitude reported in the literature [62] for nanostructured grains. This is further elaborated in Fig. 4b and 4c by polydisperse Voronoi tessellation analyses on the largest system at the level of the octahedral and tetrahedral site types (with Pd radius of 1.37 Å from bulk FCC Pd and H radius of 0.58 Å, which is the radius of a relaxed sphere inside a Pd octahedral site from this interatomic potential with 1% relative face area threshold [57,58,66] around the interstitial sites). Fig. 4b and 4c show the expansion of the bulk octahedral and tetrahedral site volume, and now we can see that at fine grain sizes of less than 8 nm, the site volumes begin to diverge higher, and thus lower the solute bulk energy in those sites. This size effect appears to be directly caused by the elastic relaxation of interstitials in the larger bulk interstices that emerge at fine grain sizes. However, it turns out that this bulk volumetric effect (which is less than 1 J/mol in Fig. 4b and 4c) is not large enough to fully explain the intrinsic size dependencies in the segregation spectra, which are >1 kJ/mol (Fig. 3b). We thus repeat the above dilation analysis for both GB and TJ average site volume and segregation energy at varying grain sizes in Fig. 5a and 5b. Unlike the bulk interstitial sites, the GB and TJ interstitial sites are remarkably affected by grain size. We see a clear and significant change in segregation energy on average, and also cf. segregation subspectral parameters in Table I., which is of the opposite trend as the bulk energies and correlates well with the site volumes: the GBs and TJs become less hospitable to solutes (cf. Fig. 5b) due to the size dependent atomic level strain. The magnitude of the energetic shift is more than twice the bulk size dependency in Fig. 4, and adds up to the expected order of 1 kJ/mol. This change in the intergranular environments is therefore the principal source of the size dependence of the segregation energy subspectra in Fig. 3b.

To gain more insight into the GB-TJ contrast of interstitial chemistry, the site volume and coordination (Z) distributions are binned and cross-plotted in Fig. 6a and 6b for GBs and TJs respectively for the d = 10.4 nm system. This mode of presentation shows an observable gradient and optimal sites for segregation, indicating that hydrogen atoms prefer sites with more free volume. Trends for coordination are not as clear; the optimal bin is near Z = 6 which is the coordination of octahedral sites, indicating that the preferential sites are structurally close to a strained bulk octahedron at a larger site volume than the bulk of closer to ~2 Å$^3$. For reference, the coordination and volume of bulk octahedral (star) and tetrahedral sites (triangle) are marked in Fig. 6a and 6b. The color scale for TJ sites also appears to be brighter; the simple Voronoi descriptors may not be able to accurately predict the contrast between GBs and TJs. To see the differences more clearly, we compare the GB and TJ interstitial sites on the volume axis individually, by plotting the site volume-energy density plot in Fig 6c indicating that while there are two distinct clusters of semi-octahedral and semi-tetrahedral sites in both GBs and TJs, the subdistributions do not overlap perfectly; there is a GB-TJ contrast that cannot be modeled with only local coordination and volume. The properties may require other atomic environment descriptors such as Smooth Overlap of Atomic Positions [67–70] or Strain Functionals descriptor [71,72], although site volume can provide a rough estimation of segregation tendency as demonstrated in Fig. 6.

TABLE II. Distribution Parameters for Segregation Energy Subspectra (in kJ/mol) for use with Eq. (1)

| d (nm) | GB | | | | | | TJ | | | | | |
|---|---|---|---|---|---|---|---|---|---|---|---|---|
| | $w_{tet}$ | $\mu_{tet}$ | $\sigma_{tet}$ | $w_{oct}$ | $\mu_{oct}$ | $\sigma_{oct}$ | $w_{tet}$ | $\mu_{tet}$ | $\sigma_{tet}$ | $w_{oct}$ | $\mu_{oct}$ | $\sigma_{oct}$ |
| 5.8 | 0.21 | 6.3 | 4.02 | 0.79 | −6.74 | 5.82 | 0.18 | 2.8 | 5.1 | 0.82 | −9.22 | 5.03 |
| 6.6 | 0.21 | 6.01 | 4.04 | 0.79 | −7.11 | 5.73 | 0.21 | 1.83 | 5.41 | 0.79 | −9.84 | 4.85 |
| 7.5 | 0.22 | 5.59 | 4.14 | 0.78 | −7.42 | 5.59 | 0.15 | 2.93 | 4.57 | 0.85 | −9.79 | 5.18 |
| 8.6 | 0.2 | 5.36 | 3.91 | 0.8 | −7.82 | 5.78 | 0.18 | 1.86 | 5.09 | 0.82 | −10.4 | 4.96 |
| 9.2 | 0.2 | 5.48 | 4.07 | 0.8 | −7.73 | 5.83 | 0.17 | 2.12 | 4.88 | 0.83 | −10.2 | 4.95 |
| 9.8 | 0.2 | 5.19 | 4.05 | 0.8 | −7.92 | 5.68 | 0.2 | 1.39 | 5.26 | 0.8 | −10.4 | 4.92 |
| 10.4 | 0.19 | 5.63 | 3.96 | 0.81 | −7.7 | 5.85 | 0.16 | 2.4 | 4.8 | 0.84 | −10.2 | 5.09 |

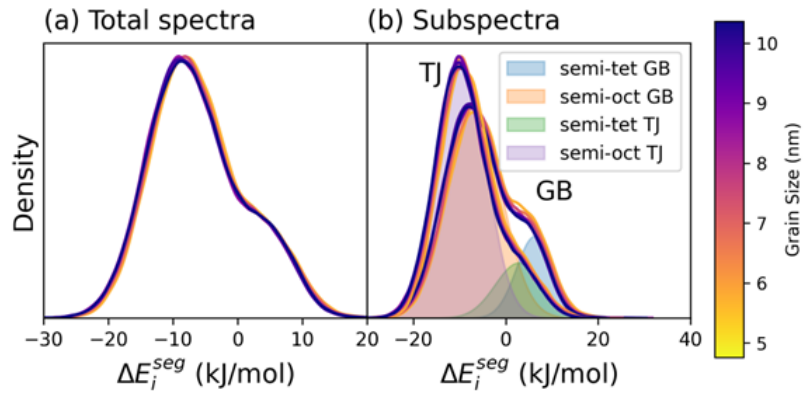

FIG. 3. (a) Segregation energy spectra at various grain sizes from the computed polycrystals. Segregation subspectra at each grain size for all grain boundaries and triple junctions calculated in (b). The fitted double-Gaussian parameters are listed in Table II.

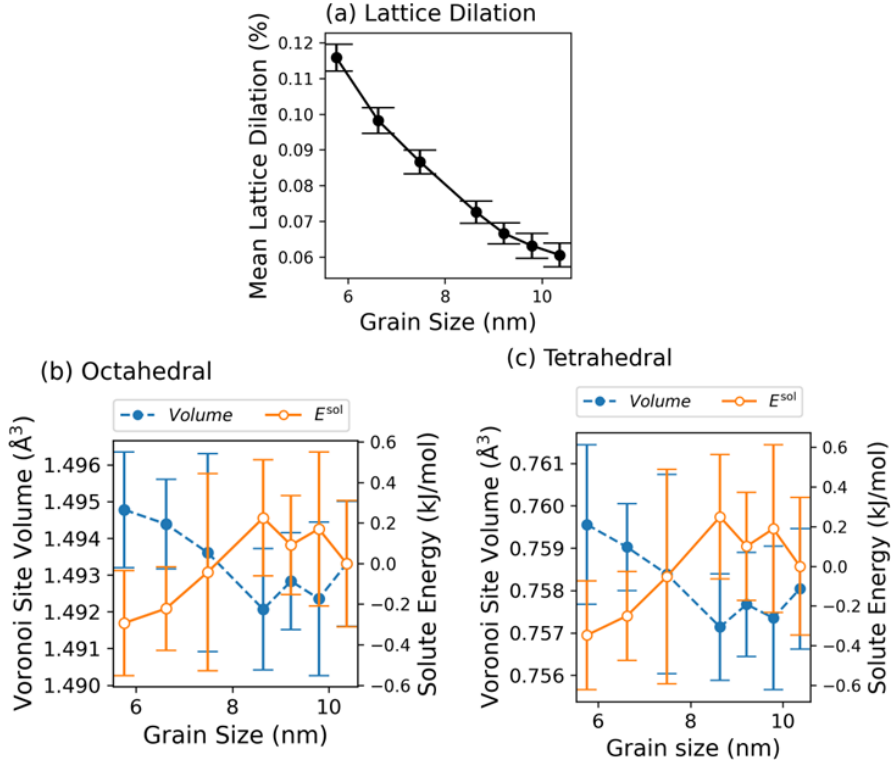

FIG. 4. Lattice dilation of Pd nanocrystals is shown in (a). The error bars indicate the standard deviation from all bulk sites at a given grain size. The sites are categorized into octahedral sites (b) and tetrahedral sites (c) which all show a drop in solute energy ($E^{sol} = E^{solute}_{bulk} - E^{poly}$) with increasing grain size. The solute energies are all referenced to the largest grain size calculated here.

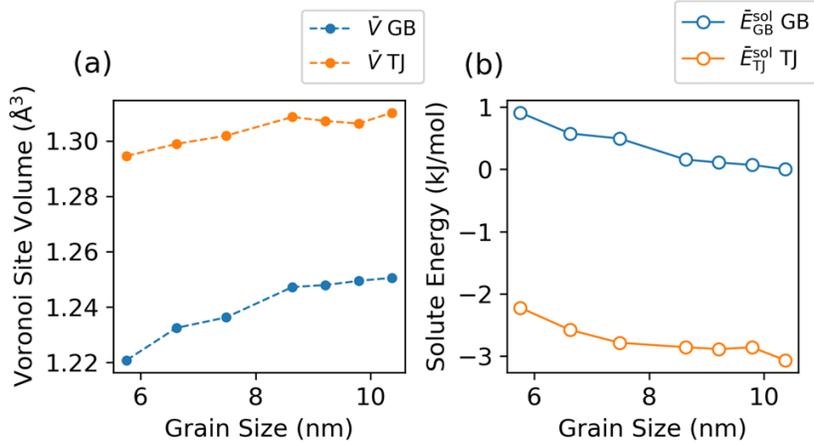

FIG. 5. The means of the (a) Voronoi site volumes and (b) site solute energy ($E^{solute}_i - E^{poly}$), referenced to the average of grain boundary sites from the largest system) are plotted here for all grain sizes calculated to show size dependencies of grain boundary and triple junction subspectra. We note that while we use averages for illustration, the distributions are multimodal (cf. Fig. 3 and Table II.).

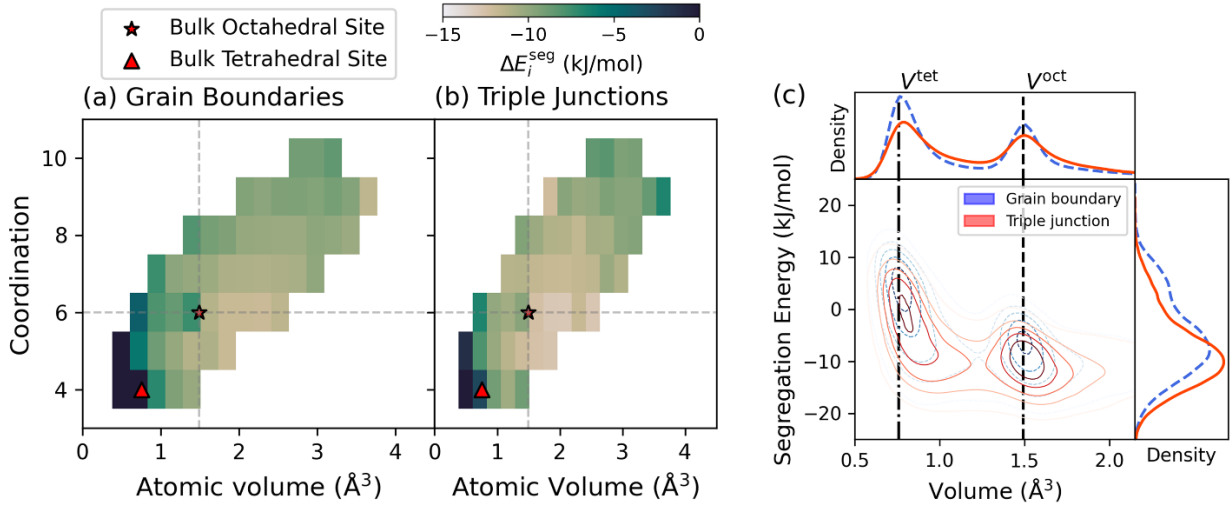

FIG. 6. Correlation between interstitial segregation energies and site volume and coordination for (a) grain boundaries and (b) triple junctions at d = 10.4 nm. The site segregation energies are averaged by binning (with a minimum sample size of 30 per bin). The density plot of both grain boundary and triple junction volume are also plotted in (c).

## V.  SIZE-DEPENDENCES OF INTERGRANULAR SOLUTE SEGREGATION

Now that we have quantified the size dependence of the segregation spectra, we can translate them through the isotherm to demonstrate equilibrium junction segregation in FCC nanocrystalline systems using Eq. (8). We first show in Fig. 7a computed isotherms as a function of total solute content for Pd(H) at fixed T = 700 K and d = 5.8 nm, based on Eq. (8). As expected, the energetic dominance of TJ sites with respect to GBs in Fig. 5 and Table I produces elevated TJ segregation vis-à-vis GBs (for example, ~5 at.% more solute at TJs vis-à-vis GBs at the global solute content of 5 at.%). This effect is expected at all grain sizes, but the way in which the individual contributions of GBs and TJs add depends on grain size through volume fractions $f^{GB}$ and $f^{TJ}$ (cf. Fig. 2 and Eq. (8)). Such local solute elevation may play a role in TJ-mediated intergranular plasticity [1,73–75], precipitation [76,77] and complexion formation [78–80], which can be driven by chemical heterogeneity in the intergranular network.

We also evaluate the size effects introduced via TJs in Pd(H) in Fig. 7b at the temperature of 700 K and $X^{tot}$ of 1 at.%. We show the equilibrium concentrations of GB, TJ and all intergranular sites as a function of grain size. As expected, the localized TJ effects are strong (also cf. Fig. 7a), but their additive contribution in the total intergranular network is generally negligible beyond a grain size of ~10 nm due to the diminishing fraction of TJ sites (or i.e. $\overline{X}^{IG}$ quickly converges to $\overline{X}^{GB}$ in Fig. 7b). However, at any grain size the local TJ concentration rises drastically due to the diminishing available sites for segregation [2]; to the extent that high concentrations of H are physically influential (e.g., in embrittlement, etc.), TJs would appear to be a prime concern at all grain sizes.

The significant solute segregation in Fig. 7 overall speaks to the bulk solubility as a whole when the intergranular site fraction ($f^{GB}$ and $f^{TJ}$) reach nonnegligible levels at the finest nanocrystalline grain sizes, which may have implications in energy applications such as hydrogen storage [81–85] or catalysis [86,87]. Here, we show a figure of merit in terms of hydrogen density per volume (in kg H per m$^3$) in Fig. 8. The hydrogen content is estimated via:

$$\rho^H = n_{i-s}(f^C X^C + f^{GB} X^{GB} + f^{TJ} X^{TJ})/V_m \tag{11}$$

where in this case we fix $X_0^C$ to be ~10 at.% and T = 700 K. The molar volume $V_m$ is estimated to be ~8.82 cm$^3$/mol from the crystalline lattice volume of pure Pd, and the number of interstitial sites per Pd atom is approximated as $n_{i-s} = 3$, although we note that those values for GBs can differ slightly [20]. The subspectral parameters are conservatively assumed to be those of the smallest grain size in Table II. due to their higher average energy vis-à-vis larger grains. With grain boundaries as hydrogen adsorbers, the bulk hydrogen density can increase more than threefold when the grain size approaches the nanocrystalline limit. The contributions from the bulk lattice, GB and TJ sites are also plotted individually, showing the significance of defects at very small grain scale of less than ~20 nm. Although this treatment misses physical complexities at higher concentrations including solute-solute interactions, structural transitions of the boundaries, and the formation of hydride products, Fig. 8 shows a large nominal effect of TJs on hydrogen storage in the finest nanocrystalline materials, one which may be worth further study and exploitation.

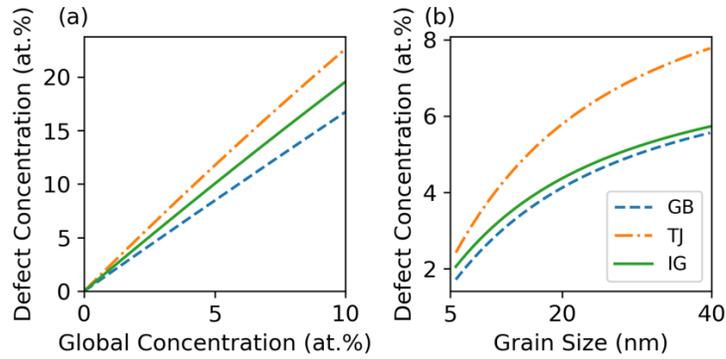

FIG. 7. Segregation concentration for grain boundary (GB dashed line) and triple junction (TJ dash-dotted line) are plotted in (a) at T = 700 K and d = 5.8 nm, and (b) as a function of grain size at fixed $X^{tot}$ of 1 at.%. Both defect types are weight averaged via their defect site fraction from Fig. 2 to produce the intergranular defect concentration (IG solid line). The segregation is calculated via Eq. (8).

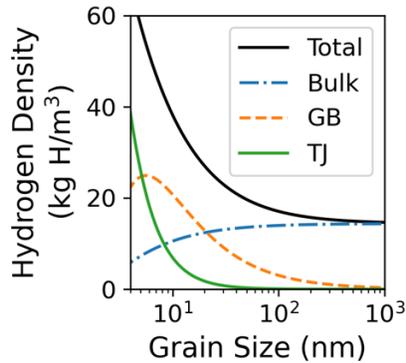

FIG. 8. Hydrogen density as a function of grain size at $X_0^C$ = 10 at.% and T = 700 K. The contributions from each site type are indicated for bulk lattice (blue dash-dotted line), grain boundary (orange dashed line) and triple junction sites (green solid line).

## VI. CONCLUSIONS

We have developed a framework to address interstitial solute segregation at finite TJ volume fractions and demonstrated applicability of the method via a model Pd(H) system. Such small grain sizes cause heterogeneity at TJs vis-à-vis GBs and size dependencies. The solute energetics at both GBs and TJs show a strong contrast, resulting in significant heterogeneity in the intergranular network. This speaks to the characteristic contrast of segregation subspectra shown in Table II. and Fig. 3, resulting in significant solute elevation at TJs (cf. Fig. 7). The total size dependencies in Fig. 3 are contributed by both the relative defect fraction from Fig. 2 and intrinsic size dependencies of GB and TJ subspectra (cf. Fig. 5). Such intrinsic size dependencies here are found to be more sensitive than substitutional segregation systems; in substitutional systems size dependencies are approximately extrinsic and caused by the relative proportions of GBs and TJs changing with grain size. In the interstitial system studied here, by contrast, the size effect is intrinsic to the site environments in the GBs and TJs being affected by dilation strain in nanocrystalline systems. In the Pd(H) system studied here, this intrinsic size effect remarkably counteracts the extrinsic shifts of the total IG spectra. This framework suggests that we may require sampling of GBs at sufficiently large system sizes for traditional alloy applications to avoid size dependencies found in this work, which are found to be significant at grain sizes below about 8 nm. Finally, the hydrogen framework may have implications in hydrogen storage, as we demonstrate that the bulk solubility of NC alloys in Fig. 8 has a multifold increase in solubility due to the dominant defect volume fractions with preferential sites for hydrogen adsorption. The work may also have implications in nanocrystalline alloy design as well, where the considerations of stabilizing TJ environments may allow pathways to overcome, for instance, thermal stability and mechanical problems in nanocrystalline materials. We look forward to seeing future work in such directions.

## ACKNOWLEDGMENTS

This work was supported by the US Department of Energy award No. DE-SC0020180 for work on the spectral framework for grain boundary segregation, and by the Office of Naval Research (ONR) under the grant N000142312004 for work on hydrogen solubility. The authors acknowledge MIT Satori and Research Computing Project for the computational resources used in this work. M. Wagih is supported by the Lawrence Fellowship at Lawrence Livermore National Laboratory. N. Tuchinda also acknowledges fruitful discussions with T. Matson at MIT.